\documentclass[twocolumn]{article}
\usepackage[utf8]{inputenc}
\usepackage[T2A]{fontenc}
\usepackage[english]{babel}
\usepackage{hyperref}
\usepackage{float}
\usepackage{indentfirst}
\usepackage{comment}
\usepackage{authblk}
\usepackage{ragged2e}
\usepackage{geometry}
\usepackage{comment}
\usepackage{lipsum}
\usepackage{subcaption}
\usepackage{array, ragged2e, mathtools}
\usepackage{tabularx}
\usepackage{wrapfig}
\usepackage{amsmath, amssymb, bm}
\usepackage{graphicx, svg}

\usepackage[title]{appendix}
\usepackage[section]{placeins}

\usepackage{titlesec}

\usepackage{dcolumn}
\usepackage{multirow}
\usepackage{rotating}
\usepackage{cite}
\usepackage{siunitx}

\title{On Frequency and Angle Stability of Anomalous Metasurface Reflectors\\ Synthesized with Auxiliary Field and Power Conformal Methods}
\author[1]{Vyacheslav Ivliev}
\author[1,*]{Stanislav Glybovski}

\affil[1]{School of Physics and Engineering, ITMO University, St.~Petersburg 197101, Russia}
\affil[*]{s.glybovski@metalab.ifmo.ru}

\date{}  

\begin{document}

\twocolumn[

\maketitle

\begin{abstract}

Metasurfaces (MSs) have recently become an efficient solution for controlling a spatial field distribution according to the demand of an electromagnetic device. In particular, anomalous reflectors based on impenetrable MSs can be realized by introducing a periodic spatial modulation of a local surface impedance defined on a flat or curvilinear surface. Two main methods to derive the profile of a purely reactive impedance that enable perfect anomalous reflection at an arbitrary angle are the power-flow conformal metasurface (PFCM) and auxiliary-field synthesis (AFS) methods. Both methods have been recently implemented with practical MS structures and have been successfully demonstrated in the experiments. However, the two methods assume drastically different synthesis algorithms and result in different impedance profiles. While the PFCM method is based on a direct analytical solution and does not require bounded wave excitation, the AFS method involves non-linear optimization of auxiliary bounded fields (surface waves) with rational selection of proper solutions. From a practical standpoint, the question of which method provides the most frequency-stabile solution becomes important to answer. Here we present a comprehensive analytical and numerical comparison of both methods applied to corrugated impedance surfaces in terms of the co-polarized anomalous reflection efficiency in a wide range of frequencies and corresponding scan angles. It is revealed that even though both methods can achieve imperceptible operational frequency bandwidths difference, the angular coverage of the frequency scan at larger reflection angles is up to 23$^{\circ}$ wider for the PFCM method.  

\end{abstract}
    \vspace{1cm} 
]

\section{Introduction}

Metasurfaces (MSs), i.e. two-dimensional arrays of discrete meta-atoms with a subwavelength step, in recent years have become an important research direction opening multiple novel applications in antennas and other electromagnetic devices. These applications in various frequency domains reviewed, e.g., in \cite{holloway2012overview, yu2014flat, glybovski2016metasurfaces} are enabled by unique possibilities of wave front transformations upon diffraction on spatially-modulated MSs, which meta-atoms are individually synthesized for every local position. The most important applications widely covered in the literature include anomalous reflectors and refractors, as well as converters from point-source radiation to an arbitrary flat wavefront (see, e.g., the works \cite{yu2011light, sun2012high, epstein2014passive, asadchy2016perfect, epstein2016synthesis, epstein2016arbitrary, diaz2017generalized, kwon2018lossless, diaz2019power, giusti2024comparison}). During the recent decade, several methods for the synthesis of MSs with these functions have been proposed, and their fundamental bounds have been explored.

Here we consider in detail the problem of co-polarized anomalous reflection, where a periodic spatial modulation of macroscopic MS parameters is to be derived and then point-wise approximated with an engineered set of discrete meta-atoms placed along the modulation period. This problem was solved in the early works using the phase gradient method adopted from reflectarray antenna design, in which the local surface impedance $Z_{\text{s}}$ is determined based on the linear reflection phase gradient assuming a unitary local reflection coefficient amplitude \cite{sun2012high}. Later, it was found \cite{asadchy2016perfect} that this method leads to a reduction in the anomalous reflection efficiency associated with spurious scattering losses, especially for high anomalous reflection angles. This issue, explained by the impedance mismatch between incident and anomalously reflected plane waves, can be addressed with several analytical design approaches. For MS implementations with local response, such as corrugated \cite{kildal1988definition} and mushroom-type \cite{sievenpiper1999high} structures for TM polarization, the approximation of a local surface impedance (Leontovitch-type boundary condition) is valid. Two main methods were proposed to derive its required spatial modulation corresponding to the perfect anomalous reflection: the power-flow conformal metasurface (PFCM) and auxiliary-field synthesis (AFS) methods. The first method consists in finding a purely imaginary $Z_{\text{s}}$ distribution and a profile function of a curvilinear MS, which sustain the desired combination of only two existing plane waves (incident and anomalously reflected), in accordance with the local power conservation principle \cite{tereshinsedovchaplin, diaz2019power}. The advantage of this method is that it provides a direct solution (without any numerical optimization involved), where the surface impedance profile can be obtained by solving an ordinary differential equation. At the same time, it necessarily results in a curved MS profile, which may be complicated to fabricate. The second method consists in finding a purely imaginary $Z_{\text{s}}$ distribution of a flat MS meeting the requirements of the local power conservation principle due to the presence of auxiliary bounded fields (surface waves) excited near the MS plane. The corresponding surface impedance profile is calculated by nonlinear optimization of the surface wave amplitudes \cite{kwon2018lossless}. Despite the benefit of a flat MS profile, multiple very different surface impedance solutions that are equally well adapted to the perfect anomalous reflection goal function can be predicted during the optimization. Disadvantageously, some of those solutions may be associated with high reactive power leading to narrowband properties. Consequently, a proper selection is required to find the solution providing the best frequency and angular stability of the anomalous reflection, as it is usually demanded in the applications. Furthermore, it is an open question how the frequency and angular properties of the best selected solution of the AFS method compare with the direct solution of the PFCM method for the same type of meta-atom employed and the same target anomalous reflection direction set. Indeed, the presence of bounded excitation waves in the AFS method may affect the bandwidth compared to that in the PFCM method, in which the total field outside of the structure consists of only two propagating plane waves.

It should be noted that the efficiency of analytically synthesized anomalous reflectors may further deteriorate due to the effect of spatial dispersion (nonlocality) in some structures, such as metal patch grids on a grounded dielectric slab \cite{movahediqomi2023comparison, yepes2022role}. In this case, the MS cannot be modeled using local surface impedance, so both PFCM and AFS methods become inaccurate. However, for that particular structure, instead, a local grid impedance could be modeled along with a nonlocal grounded dielectric slab, which improves the anomalous reflection efficiency at high angles \cite{movahediqomi2023comparison}.

From a practical point of view, local surface impedance boundary conditions can be realized with several different geometries. In the short-centimeter, millimeter wave and terahertz ranges, which are important for prospective wireless communication systems, one of the promising implementations of spatially modulated impenetrable MSs is based on corrugations with a subwavelength step made in metal plates via precise micromachining \cite{cai2016leaky} or 3D printing \cite{beaskoetxea20163}. Moreover, corrugated surfaces due to the near-quarter-wavelength depth of grooves typically provide a larger bandwidth compared to printed-circuit-board (PCB) periodic structures having an electrically small thickness, at least when both technologies are applied to reflectarray antennas \cite{Sangster2019}. 

Therefore, the operational bandwidth of an anomalous reflector, or other low-profile beam-shaping device based on an analytically synthesized surface impedance spatial modulation, depends both on the method of synthesis and the meta-atom implementation. For practical applications, especially, for relatively broadband wireless communication systems, the crucial question is how to achieve the largest possible bandwidth. In fact, most of the available MS synthesis methods and the corresponding results are available in the literature only for one frequency, or are limited with observing narrow frequency bands. Except for the comparison of different synthesis results for thin grounded metal patch grids with non-local properties discussed in \cite{movahediqomi2023comparison}, to the best of our knowledge, no systematic comparison has been performed aimed at revealing the optimal synthesis approach for anomalous reflectors in terms of bandwidth.

In this work, we analytically and numerically compare the frequency properties of solutions obtained for co-polar anomalous reflectors based on spatial modulation of local surface impedance via the main two methods, i.e., the PFCM and AFS methods. With this aim, we assess the efficiency of the reflected beam in a wide frequency range. The two methods are compared for the same anomalous reflection directions required at the frequency of interest, and implemented in numerical simulations using the 1-D corrugated metal structure providing a reasonable compactness-bandwidth compromise.

This paper is organized as follows. In Section \ref{sec:methods} the analytical approaches of both the PFCM and AFS methods are presented and compared, the frequency dispersion model of the corrugated structure is summarized, and the approach to selection of the nonlinear optimization results of the AFS method is discussed. In Section \ref{sec:results}, corrugated structures synthesized with both methods for the same anomalous reflection angles are numerically evaluated in terms of frequency and angular stability.

\section{\label{sec:methods}Analytical methods}

\begin{figure*}[h!]
\includegraphics{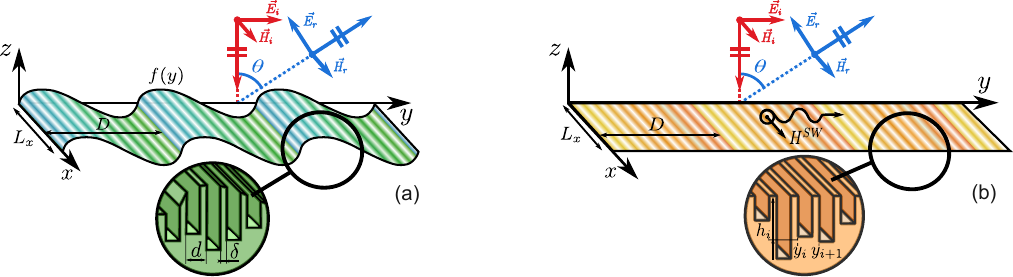}
\centering
\caption{\label{Fig_1_Problem_of_anomalous_reflection}Geometry of the problem with a TM-polarized plane wave anomalously reflected from a power-flow conformal (a) and flat (b) impenetrable impedance MS with a modulation period of $D$ along the $y$-axis. Each MS is considered as a perfectly conducting corrugated structure with a step of grooves equal to $d$ as shown in the insets.}
\end{figure*}

We consider anomalous reflection of a TM-polarized plane wave that normally impinges on an MS from the positive $z$-axis direction. The MS is periodically modulated in the $y$ direction. While for the PFCM method (Fig.~\ref{Fig_1_Problem_of_anomalous_reflection}(a)) the MS is curved according to a certain periodic profile function $z=f(y)$, for the AFS method (Fig.~\ref{Fig_1_Problem_of_anomalous_reflection}(b)) the MS is flat and occupies the $xy$-plane. Both problems are assumed two-dimensional (no variation along the $x$ direction). MS impedance is modulated with period 
\begin{equation}
    D = \dfrac{\lambda}{|\sin \theta|} 
    \label{Modulation_period}
\end{equation} with $\lambda$ being the free space wavelength, in order to create the first propagating  higher-order Floquet–Bloch (FB) mode outgoing from the MS at the desired anomalous reflection angle $\theta$ in the $yz$-plane. A monochromatic time dependency $e^{j \omega t}$, where $\omega$ is the angular frequency, is assumed throughout this paper.

For the TM-polarization the incident and desired reflected waves 
field components can be written as 
\begin{align}
\begin{split}
    \mathbf{E}_i = E^i_y \mathbf{e}_y,&  \quad \mathbf{H}_i = H^i_x  \mathbf{e}_x \\
    \mathbf{E}_r = E^r_y  \mathbf{e}_y + E^r_z  \mathbf{e}_z,& \quad   \mathbf{H}_r = H^r_x  \mathbf{e}_x, 
    \label{Desired_fields}
\end{split}
\end{align} where the relation between their wave amplitudes can be derived from the integral energy conservation law \cite{asadchy2016perfect, mohammadi2016wave}:
\begin{equation*}
    \dfrac{|E^r_y|}{\cos \theta} = \dfrac{|E^r_z|}{\sin \theta} = \dfrac{|E^i_y|}{\sqrt{\cos \theta}}.
\end{equation*} 
At the same time, the initial phase $\varphi_{\text{r}}$ of the reflected wave (i.e. the phase that the $y$-component of the reflected electric field has at the origin) can be set arbitrarily.

In both methods, the spatial modulation of the purely reactive surface impedance is obtained based on the same requirement of real part of the Poynting vector's normal component to be equal to zero at each point of a MS (local passivity requirement coming from energy conservation law):
\begin{equation}
    \mathfrak{Re} \Big( S^+_n(y) \Big) = \frac{1}{2} \mathfrak{Re} \Big( E^+_{\tau}(y) \, H^+_{\tau}(y)^* \Big) = 0,
    \label{local_passivity_requirement}
\end{equation}
where $E^+_{\tau}(y)$ is the tangential component of total electric field (the same notation is used for magnetic field $H^+_{\tau}(y)$) just above the MS and $S^+_n(y)$ is the normal component of the Poynting vector. However, the approaches to define $E_{\tau}^{+}(y)$ and $H_{\tau}^{+}(y)$ for both methods are different, as discussed in the following subsections.

\subsection{Power-flow conformal metasurface method}

The aim of this method is to find a power-flow conformal metasurface (PFCM) profile function $f(y)$ and impedance distribution $Z(y,f(y))$ along it which ensure the real power crossing the MS to be locally conserved for the total field $E^+_{\tau}(y) \triangleq E_{\tau}(y, z) |_{z \to f(y)^+}$ and $H^+_{\tau}(y) \triangleq H_{\tau}(y, z) |_{z \to f(y)^+}$ being a sum of the incident wave and the desired anomalously reflected plane wave \cite{tereshinsedovchaplin, diaz2019power}. The condition \eqref{local_passivity_requirement} is equivalent to
\begin{equation}
    \mathfrak{Re} \, Z(y, f(y)) = 0.
    \label{ReZ=0_PFCM}
\end{equation}

According to \cite{tereshinsedovchaplin} the condition \eqref{ReZ=0_PFCM} leads to the differential equation containing function $f(y)$:
\begin{equation}
    \label{f'(y)_diff_eq_PFCM}
    f'(y) = \dfrac{-\sqrt{\cos \theta} \, \operatorname{tg}  \dfrac{\theta}{2}  \, \cos  \beta(y)  }{ 1 + \sqrt{\cos \theta} \, \cos  \beta(y) },
\end{equation}
where $\beta(y) = k \, f(y) \, (1+ \cos\theta ) + ky \sin\theta - \varphi_r$ and $k~=~2\pi/\lambda$ is the free space wavenumber.

One can show that \eqref{f'(y)_diff_eq_PFCM} can be be integrated in quadratures to obtain a general solution in the form of an implicitly defined function \cite{diaz2019power}
\begin{equation}
    \label{General_solution_f_y_implicitly}
    f(y) + \dfrac{\sqrt{\cos\theta} \, \operatorname{tg}  \dfrac{\theta}{2} }{k \sin\theta} \, \sin  \beta(y) = C,
\end{equation}
where $C$ is an arbitrary constant defining a variety of possible profiles. Once the profile function is obtained, the purely imaginary impedance distribution along the PFCM is determined by \cite{tereshinsedovchaplin, diaz2019power}
\begin{equation}
    Z(y, f(y)) = \dfrac{ j Z_0  \sqrt{\cos \theta} \, \sin  \beta(y) }{\sqrt{1 +\big(f'\left(y\right)\big)^2 } \Big( 1 + \sqrt{\cos \theta} \, \cos \beta(y) \Big)},
    \label{Z(y,f(y))_PFCM}
\end{equation}
where $Z_0$ is the wave impedance of free space.

\begin{figure}
\centering
\includegraphics{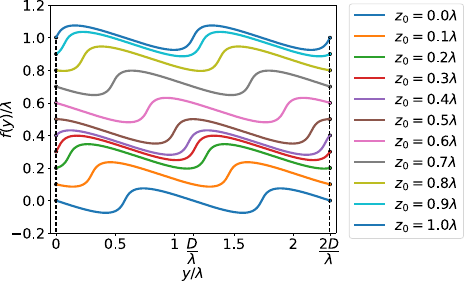}
\caption{The family of curved MS profiles with a modulation period of $D$ and periodic impedance distribution along it for different starting points with $y_0=0$ and various $z_0$ for $\theta=60^{\circ}$ and $\varphi_{\text{r}}=0$.}
\label{Fig_2_PFCM_profiles_for_different_starting_point}
\end{figure}

For the given incident and reflected plane waves \eqref{General_solution_f_y_implicitly} defines a family of non-flat curves $f(y; z_0)$ all having period $D$ and periodic impedance distribution $Z(y, f(y); z_0)$ along it. Both functions depend on the parameter $z_0$, which is a vertical coordinate of the starting point of integration ($y_0$, $z_0$) in \eqref{f'(y)_diff_eq_PFCM} or, equivalently, on the choice of the constant $C$ in \eqref{General_solution_f_y_implicitly}. From the analysis of \eqref{f'(y)_diff_eq_PFCM} it can be shown that the choice of different starting points only results in the translational shift of the same profile due to the periodicity of the right-hand side in \eqref{f'(y)_diff_eq_PFCM}.

In Fig.~\ref{Fig_2_PFCM_profiles_for_different_starting_point} a family of profile curves calculated for $\theta=60^{\circ}$ and $\varphi_r=0$ is depicted as an example. As can be seen, all periodic profiles that create the given reflected wave differ only by a translational shift. From the periodicity of the right-hand part of \eqref{Z(y,f(y))_PFCM}, it can further be demonstrated that all periodic profiles are associated with the same purely imaginary impedance distribution within a period, which is given directly by \eqref{Z(y,f(y))_PFCM}. Moreover, it can be shown that the choice of $\varphi_r$ also leads to profile curves being distinguished from each other by only a translational shift, all having the same impedance distribution. In other words, the solution in the PFCM method is unique for the given value of $\varphi_r$ at the arbitrary starting point with a vertical position of $z_0$. For this reason, without loss of generality, in the following we consider the profile obtained for $y_0=0,z_0=0$ and $\varphi_r=0$ as frequency properties for all other profiles of the family are similar. Note that in this method, nonlinear optimization is not required to calculate the macroscopic parameters of the structure.

\subsection{Auxiliary-field synthesis method}

As the MS becomes flat and occupies the plane $z=0$, the presence of only incident and reflected plane waves leads to the presence of nonzero real and imaginary parts of $Z$ simultaneously, thus violating the local passivity condition (however, the MS remains overall passive as the period-averaged value of the real part is zero) \cite{mohammadi2016wave, asadchy2016perfect}. To satisfy condition \eqref{ReZ=0_PFCM} for the case of a flat MS ($f(y)\equiv 0$) it is necessary to introduce auxiliary fields (that is, to accept excitation TM-polarized surface waves) \cite{epstein2016synthesis, kwon2018lossless} propagating along the $y$ axis. With the surface wave fields $E_{y}^{\text{sw}}(y)$, $H_{x}^{\text{sw}}(y)$ added the total tangential field components should relate through a purely imaginary surface impedance, i.e.:
\begin{equation}
    \mathfrak{Re} Z (y) = \mathfrak{Re} \dfrac{E^i_y (y) + E^r_y(y) + E_{y}^{\text{sw}}(y)}{H^i_x(y) + H^r_x(y) + H_{x}^{\text{sw}}(y)} = 0.
    \label{ReZ=0_AFS}
\end{equation}

Providing the above conditions implies finding unknown periodic distributions of $E^{\text{sw}}, H^{\text{sw}}$. Since the MS is periodically modulated, a mathematical apparatus of FB modes can be applied. The tangential field components just above the MS can be expanded in Floquet series as
\begin{align}
\begin{split}
    E_{y}(y) &= \sum\limits_{n = - 1 - N_{l}}^{1 + N_{r}} E_{n} \, e^{-j q_n y}, \\
    H_{x}(y) &= \sum\limits_{n = - 1 - N_{l}}^{1 + N_{r}} H_{n} \, e^{-jq_n y}, 
    \label{E_H_Floquet_expansion}
\end{split}
\end{align} where $q_n = 2 \pi n/D$, while $E_n$ and $H_n$ are yet unknown amplitudes of FB modes connected to each other for each number $n$ with Maxwell equations. 

For the means of surface-wave optimization in \eqref{E_H_Floquet_expansion} we center the FB-mode basis with respect to the incident wave (i.e. $n=0$ corresponds to normal incidence), identifying the desired anomalously reflected wave with $n=1$. At the same time, the undesirable mirror-symmetric mode with $n=-1$ should have zero amplitude, whereas higher-order  (evanescent for $\lambda < D < 2\lambda$) modes with $|n|>1$ take the amplitudes to be determined. Note that in \eqref{E_H_Floquet_expansion} we cut the series to limit the number of unknowns during optimization. Unlike \cite{kwon2018lossless} here we consider not only $N_r$ FB modes ($n>1$) propagating in the positive direction of the $y$ axis (the same as the direction of the anomalous reflection), but also $N_l$ FB modes ($n<-1$) propagating in the negative direction. Thereby, to accomplish \eqref{ReZ=0_AFS} in the sense of pointwise convergence at the chosen set of evenly spaced nodes $\Big\{y_i \Big\}_{i=0}^{2 \cdot (N_r + N_l) - 1}$ within the modulation period, we numerically search for $2 (N_r + N_l)$ unknowns (real and imaginary parts of $E_n$) that satisfy the nonlinear equations:
\begin{equation}
    \Big\{  \mathfrak{Re} S^+_z(y_i) = 0 \Big\}_{i = 0}^{{2 (N_r + N_l)} - 1}.
    \label{total_eq_syst_AFS}
\end{equation}
With this aim the \textit{fsolve} function in Python based on a modification of the Powell hybrid method and implemented in MINPACK \cite{more1980user} is employed. Indeed, the convergence rate in this approach turns out to be higher for the considered equation system than in other numerical methods. It should be noted that, in contrast to \cite{kwon2018lossless} $N_r$ and $N_l$ are considered two independent parameters. Negative indices of the FB modes have also been considered in \cite{giusti2024comparison} for the problem of anomalous refraction with an equal number of positive and negative evanescent modes. However, as shown in the following, $N_r$ and $N_l$ have different effects on the resulting impedance distribution.

\subsection{Frequency-dependent surface impedance of a corrugated structure}

Both solutions discussed above for the surface impedance profile precisely enable anomalous reflection into the given angle $\theta$ only at one frequency $f_0$ of interest. There are three main effects that deteriorate the operation of the synthesized MS anomalous reflector at $f \neq f_0$, as follows. The first effect is that the solutions given by the expressions \eqref{ReZ=0_PFCM} or \eqref{ReZ=0_AFS} are valid only at $f=f_0$. As the frequency shifts, spurious reflection to undesirable propagating FB modes leads to a reduction in reflection efficiency, i.e. the reduction in the magnitude of the desired FB mode with an index of $n=1$. The second effect consists in the frequency scan of the anomalously reflected beam according to \eqref{Modulation_period}. Finally, the third effect is related to the unavoidable frequency dispersion of a purely imaginary surface impedance at every point of the synthesized profile according to Foster's theorem. The law of frequency dispersion, in turn, depends on the microstructure of meta-atoms.

Here, we aim to compare the PFCM and AFS methods in terms of the best achievable frequency stability of the anomalously reflected beam. In fact, we only compare the influence of the first effect in both methods, as the second one is common for any anomalous reflector with a given modulation period. To minimize the third effect, we consider metal corrugated structures that implement the synthesized local surface impedance profiles providing a generally larger bandwidth in comparison to thin printed structures \cite{Sangster2019}. A meta-atom of a uniform periodic corrugated structure having a subwavelength step of $d$ in the $y$ direction contains a groove of depth $h$ separated from the neighboring grooves by metal walls of thickness $\delta$ (see insets in Fig.~\ref{Fig_1_Problem_of_anomalous_reflection}(a) and Fig.~\ref{Fig_1_Problem_of_anomalous_reflection}(b)). In the following, an equivalent-circuit model of the meta-atom employed to synthesize the microstructure of a corrugated anomalous reflector is discussed in brief. 

\begin{figure}
\centering
\includegraphics[width=0.9\columnwidth]{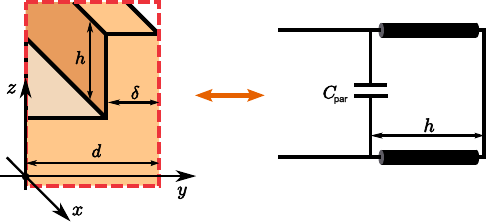}
\caption{The geometry (left) and equivalent circuit (right) of a single corrugation with a depth $h$ cut in perfectly conductive metal within a meta-atom with a sub-wavelength size $d \ll \lambda$.}
\label{Fig_3_Equivalent_groove_scheme}
\end{figure}

The parameters of each corrugation can be predicted using a very well-known transmission line model \cite{kildal1988definition, Markov1979electrodynamics}. An equivalent circuit of a single corrugation contains an air-filled transmission line section of wave impedance $Z_w=Z_0$ and length $h$ with a short-circuit load at the end (see Fig.~\ref{Fig_3_Equivalent_groove_scheme}). In order to account for the fringing field effect in the aperture of the corrugation, it is necessary to introduce the parasitic capacitance $C_{\text{par}}$. The input impedance $Z_{\text{in}}$ can be found as a parallel connection of the short-circuited line input impedance $Z_{\text{line}}=j Z_0 \operatorname{tg} \left( \omega h/c \right)$ (see, e.g., \cite{kildal1988definition}; $c$ is free space wave velocity) and the parasitic capacitive impedance $Z_C = 1 / (j \omega C_{\text{par}})$.

Then the surface impedance of the uniform periodic corrugated structure can be found as
\begin{equation}
    \label{Z_in_single_groove}
     Z = \dfrac{Z_{\text{line}} \, Z_C}{Z_{\text{line}} + Z_C} \left(1 - \dfrac{\delta}{d} \right),
\end{equation} where the coefficient $(1-\delta/d)$ takes into account the presence of metal walls between corrugations when averaging the tangential electric and magnetic fields across the upper plane of the structure.
    
For the PFCM method, the surface impedance profile obtained is discretized into 20 meta-atoms within a modulation period $D$. Each meta-atom is implemented as a corrugation with an analytically calculated depth and aperture height. With this aim, a set of 21 equidistant nodes with coordinates $\{ y_i \}_{i=1}^{21}$ with step $d$ and $\{ z_i \}_{i=1}^{21}=f(y_i)$ is selected and the exact impedance value $Z(y_i, f(y_i); f_0)$ is calculated at $f=f_0$ for each node. 
Next, the calculated profile of the PFCM $f(y)$ is approximated by a piecewise linear function of 20 sections (for the $i$-th meta-atom, the profile function is approximated with a line between the $i$-th and $(i+1)$-th nodes). 
The corrugation depth $h_i$ of the $i$-th meta-atom is calculated from \eqref{Z_in_single_groove} to provide the impedance $Z_i(f_0)$, which is taken as the arithmetic mean between exact impedances at two nodes bounding the corresponding meta-atom, i.e.:
\begin{equation}
    Z_i = j X_i = \dfrac{Z(y_i, f(y_i)) + Z(y_{i+1}, f(y_{i+1}))}{2}.
    \label{Z_i(f_0)_groove}
\end{equation}
The depth is measured from the middle point of the corresponding linear section. 

The implementation of the impedance profile obtained with the AFS method is similar to the one discussed above except that the surface profile function $f(y)$ is zero at each node, and the depths of the corrugations are measured from the $xy$-plane. 

In both methods, the corrugations occupy the left ($1 - \delta/d=2/3$) part of the corresponding meta-atoms. The value of $C_{\text{par}}$ to be used when predicting the values of $h_i$ in both methods with \eqref{Z_in_single_groove}, should first be extracted numerically. With this aim, the surface impedance of a uniform flat periodic structure (composed of identical corrugations) is calculated in CST Microwave Studio under normal incidence and compared with its value predicted by \eqref{Z_in_single_groove} for arbitrary $0 < h \leq \lambda_0/4$.

\subsection{Bandwidth comparison and solution selection criteria}

As mentioned above, while the impedance distribution and profile functions obtained using the PFCM for $f=f_0$ come from the direct solution, the AFS method relies on nonlinear optimization. Therefore, with the AFS method depending on the composition of excitation surface waves one may obtain many different impedance functions leading to almost the same anomalous reflection efficiency, especially when considered in a wide range of frequencies. To compare the achievable operational frequency bandwidths of both methods, one should select the best solution from a variety of solutions synthesized for $f=f_0$ by the AFS method and compare it with the direct solution of the PFCM method. Note that from electromagnetic standpoint, the true solution of the problem is unique, but any achievable solution of the AFS method for a limited number of surface waves $N_l$, $N_r$ is only an approximation.

In general, for $1<D/\lambda < 2$ the MS can create three propagating FB modes with indices $1$ (the desired anomalous beam), 0 (the undesired broadside beam), and $-1$ (the undesired beam symmetric with respect to the desired one) with the corresponding scattering parameters $S_1$, $S_0$, and $S_{-1}$. An ideal anomalous reflector is characterized by $S_{-1}=S_0=0$, and $S_1=1$. In the results section, we define the operational frequency band $\Delta f$ relative to the central frequency $f_0$ so that $|S_1|$ stays above the level of $0.9$. It is worth mentioning that the frequency scan effect is ignored in this definition (only the level of the desired beam is important). 

The proposed method to select the best AFS method solution is as follows. First, for the given $\theta$ we obtain a set of solutions optimized at $f=f_0$ for different combinations of $N_l$ and $N_r$ in \eqref{E_H_Floquet_expansion}. Second, we select solutions leading to good enough operation of the anomalous reflector at $f=f_0$, i.e. the solutions that exceed the direct solution by the PFCM method in terms of \textit{wave front purity} $\zeta$ defined as
\begin{equation}
    \zeta = \Big| S_1 \Big|^2 / \Big( \Big| S_0 \Big|^2 + \Big| S_{-1} \Big|^2 \Big).
    \label{Numerical_criterion_for_purity}
\end{equation} 
Note that theoretically for the PFCM method the value of $\zeta$ approaches infinity. However, in practice, once discretization is applied to design a periodic structure, spurious scattering appears and $\zeta$ becomes finite. For example, with the parameters selected above ($\theta=60^{\circ}$ and $D/d=20$) the purity of the wave front decreases to $61.3$, which was calculated numerically. Therefore, to determine proper $N_l$ and $N_r$ as well as the surface wave excitation amplitudes in the AFS method, only solutions with $\zeta > 61.3$ should be kept. Third, among the selected solutions of the AFS method with sufficiently high wave front purity, the best is determined in terms of $\Delta f/f_0$, and finally compared to the bandwidth of the PFCM method solution. All solutions are compared for the same discretization.

To compare the synthesis results, both methods are applied to build all-metal corrugated reflectors that provide seven different anomalous reflection angles ranging from $32^{\circ}$ to $85^{\circ}$ with $\varphi_r=0^{\circ}$. In the case of the PFCM method, the constant $C$ in \eqref{General_solution_f_y_implicitly} is chosen equal to $0$ corresponding to the MS curve $f(y)$ starting from the point $(0,0)$. In the case of the AFS method, we consequently solve the nonlinear optimization problem \eqref{total_eq_syst_AFS} for $N_l$ that ranges from 0 to 15 while keeping $N_r=40$ (16 solutions). Note that such a large enough number $N_r$ provides enough degrees of freedom in the numerical solution of \eqref{total_eq_syst_AFS} for each $N_l$ from the range. However, $N_l$ has a strong effect on the impedance profile obtained and the frequency behavior of the reflector. Hence, for each $N_l$, we perform a numerical calculation of the S-parameter spectra with CST Microwave Studio and find $\Delta f$. Then the realization with the highest $\Delta f$ is selected taking into account the criterion based on \eqref{Numerical_criterion_for_purity} to compare it with the result of the PFCM method for the same reflection angle. Note that the frequency spectra of the scattering parameters and the purity at $f_0$ are calculated by modeling the entire modulation period $D$ of a perfectly conducting corrugated structure with 20 meta-atoms.

\section{\label{sec:results}Results}

\begin{table*}[h!]
\caption{\label{Table_Groove_heights_theta_60_parameters} The geometric parameters~(in mm) of 20 meta-atoms within one modulation period $D$ for anomalllous reflectors synthesized using PFCM and AFS methods for $\theta=60^{\circ}$, $f_0=10$~GHz, and $\varphi_r=0$.}
\begin{tabular}{p{0.1cm} p{0.1cm}>{\centering\arraybackslash\ensuremath}p{0.45cm} >{\centering\arraybackslash\ensuremath}p{0.45cm} >{\centering\arraybackslash\ensuremath}p{0.45cm} >{\centering\arraybackslash\ensuremath}p{0.45cm} >{\centering\arraybackslash\ensuremath}p{0.45cm} >{\centering\arraybackslash\ensuremath}p{0.45cm} >{\centering\arraybackslash\ensuremath}p{0.45cm} >{\centering\arraybackslash\ensuremath}p{0.45cm} >{\centering\arraybackslash\ensuremath}p{0.45cm} >{\centering\arraybackslash\ensuremath}p{0.45cm} >{\centering\arraybackslash\ensuremath}p{0.45cm} >{\centering\arraybackslash\ensuremath}p{0.45cm} >{\centering\arraybackslash\ensuremath}p{0.45cm} >{\centering\arraybackslash\ensuremath}p{0.45cm} >{\centering\arraybackslash\ensuremath}p{0.45cm} >{\centering\arraybackslash\ensuremath}p{0.45cm} >{\centering\arraybackslash\ensuremath}p{0.45cm} >{\centering\arraybackslash\ensuremath}p{0.45cm} >{\centering\arraybackslash\ensuremath}p{0.45cm} >{\centering\arraybackslash\ensuremath}p{0.45cm}}
\hline
\hline
\multicolumn{2}{c}{$i$} & $1$ & $2$ & $3$ & $4$ & $5$ & $6$ & $7$ & $8$ & $9$ & $10$ & $11$ & $12$ & $13$ & $14$ & $15$ & $16$ & $17$ & $18$ & $19$ & $20$ \\
\hline
\multirow{3}*{\begin{turn}{90}PFCM\end{turn}} & $y_i$ & $0$ & $1.7$ & $3.5$ & $5.2$ & $6.9$ & $8.7$ & $10.4$ & $12.1$ & $13.9$ & $15.6$ & $17.3$ & $19.1$ & $20.8$ & $22.5$ & $24.3$ & $26.0$ & $27.7$ & $29.4$ & $31.2$ & $32.9$ \\
%\cline{2-22}
%\vspace{1pt}
 & $z_i$ & $0$ & $\mathllap{-}0.4$ & $\mathllap{-}0.8$ & $\mathllap{-}1.2$ & $\mathllap{-}1.6$ & $\mathllap{-}1.9$ & $\mathllap{-}2.2$ & $\mathllap{-}2.4$ & $\mathllap{-}2.5$ & $\mathllap{-}2.4$ & $\mathllap{-}1.9$ & $\mathllap{-}0.5$ & $1.8$ & $2.2$ & $2.1$ & $1.9$ & $1.7$ & $1.3$ & $0.9$ & $0.5$ \\
%\cline{2-22}
%\vspace{1pt}
 & $h_i$ & $0.3$ & $0.8$ & $1.3$ & $1.8$ & $2.3$ & $2.9$ & $3.4$ & $3.8$ & $4.2$ & $4.2$ & $1.7$ & $11.7$ & $10.8$ & $11.4$ & $12.0$ & $12.6$ & $13.2$ & $13.8$ & $14.3$ & $14.8$ \\
\hline
%\vspace{1pt}

\multirow{2}{*}[\dimexpr-0.5ex]{\begin{turn}{90}AFS\end{turn}} & $y_i$ & $0$ & $1.7$ & $3.5$ & $5.2$ & $6.9$ & $8.7$ & $10.4$ & $12.1$ & $13.9$ & $15.6$ & $17.3$ & $19.1$ & $20.8$ & $22.5$ & $24.3$ & $26.0$ & $27.7$ & $29.4$ & $31.2$ & $32.9$ \\
%\cline{2-22}
%\vspace{1pt}
% & $z_i$ & $0$ & $0$ & $0$ & $0$ & $0$ & $0$ & $0$ & $0$ & $0$ & $0$ & $0$ & $0$ & $0$ & $0$ & $0$ & $0$ & $0$ & $0$ & $0$ & $0$ \\
%\cline{2-22}
%\vspace{1pt}
 & $h_i$ & $13.7$ & $1.8$ & $2.6$ & $2.5$ & $4.5$ & $4.5$ & $5.1$ & $5.3$ & $8.5$ & $7.8$ & $8.0$ & $8.6$ & $8.7$ & $8.8$ & $9.7$ & $11.4$ & $11.1$ & $11.4$ & $14.7$ & $0.3$ \\
\hline
\hline
\end{tabular}
\end{table*}

\begin{figure*}[h]
\includegraphics{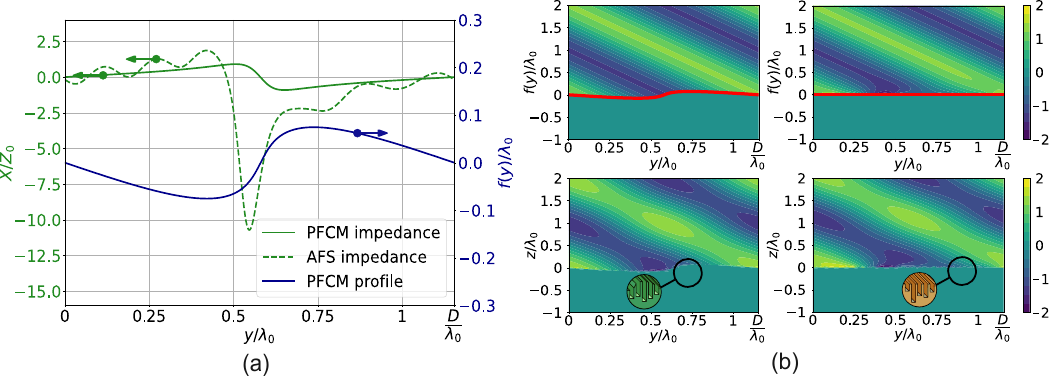}
\centering
\caption{\label{Fig_4_Syntesis_examples}Synthesis examples obtained using the PFCM and AFS methods for $\theta=60^{\circ}$: (a) MS profile for PFCM method and impedance profiles for both methods; (b) spatial distributions of the real part of the normalized E-field $z$-component $\Re (E_z/|E^i_y|)$ (left column corresponds to PFCM, right -- to AFS; the first row shows analytically predicted field distributions, the second -- realized in CST Microwave Studio for the corrugated structures with $d=1.7$~mm, $\delta=0.6$~mm, and $D=34.6$~mm, designed for $f_0=10$~GHz with geometric parameters feasible for manufacturing via micromachining.}
\end{figure*}

In the following, our comparison is shown for the operational frequency $f_0 = 10$~GHz ($\lambda_0=30$~mm). Let us first consider an example of $\theta=60^{\circ}$. In this case $D=34.6$~mm, $d=D/20=1.7$~mm, and $\delta=d/3=0.6$~mm. Note that the dimensions of the corrugations allow for the use of the smallest cutter diameter of 0.5 mm when producing the corrugated structure via micromachining. In this case, the numerically extracted value of $C_{\text{par}}$ is found to be $1.59 \cdot 10^{-15}$~ F, which is almost independent of the corrugation depth (with the exception of $h \approx 0, \lambda/4$). Note that the law of frequency dispersion of the surface impedance predicted by \eqref{Z_in_single_groove} with the extracted value of $C_{\text{par}}$ was numerically verified in the frequency range of 5 to 15 GHz. The analytically predicted values of $h_i$ that implement the required impedance profile calculated using both methods are listed in Table~\eqref{Table_Groove_heights_theta_60_parameters}. The model of the entire modulation period of the reflector is confined to the Floquet cell with dimensions $20~\text{mm}\times D \times 16.7~\text{mm}$ for the PFCM method and $20~\text{mm}\times D \times 14.3~\text{mm}$ for the AFS method. The dimensions are indicated along the $x$, $y$ and $z$ axes, respectively, with periodic boundary conditions set for the $x$ and $y$ directions. 

Examples of synthesis results for $\theta = 60^{\circ}$ (that is, the surface impedance distribution for both methods and the MS profile for the PFCM method) are shown in Fig.~\ref{Fig_4_Syntesis_examples}(a). The displayed solution for the AFS method for $N_r = 40, N_l=7$ is the one selected according to the criterion discussed above. 

For $\theta=60^{\circ}$, it turns out that only 6 solutions out of 16 satisfy the criterion, and the maximum relative operational bandwidth for the selected solution is $47.0\%$. It should be noted that without selection based on \eqref{Numerical_criterion_for_purity}, the maximum bandwidth reaches $51.8\%$, which corresponds to $N_l=8$, leading, however, to larger spurious scattering for the same discretization, compared to the PFCM method. 

The distributions of the field component $E_z$ analytically predicted with \eqref{Desired_fields} (first row) and numerically calculated for the corresponding corrugated structures in CST Microwave Studio (second row) are presented in Fig.~\ref{Fig_4_Syntesis_examples}(b). In contrast to the PFCM method, where the reflected field is expected to be just one TM-polarized reflected plane wave, for the AFS method the presence of quite strong surface waves is expected. As can be seen, this difference can be recognized in the numerical field maps related to the synthesized corrugated reflectors. However, for both methods, the reflected field has some residual distortion (the amount of power of spurious propagating FB modes is $1.3\%$ for the AFS method and $1.6\%$ for the PFCM method). As can be shown numerically, the observed distortion results from discretization into finite-sized meta-atoms. For example, discretization into 20 meta-atoms per period $D$ in the PFCM method (providing feasible dimensions of corrugations at 10~GHz using micromachining methods) provides the wave front purity of only 61.3, while for 100 meta-atoms per period, the purity increases significantly to 508.0. That corresponds to relatively narrow corrugations (not feasible for micromachining), but for the power captured by spurious propagating FB modes reduced to $0.2\%$.  

\begin{figure}[h]
\centering
\includegraphics[width=0.48\textwidth]{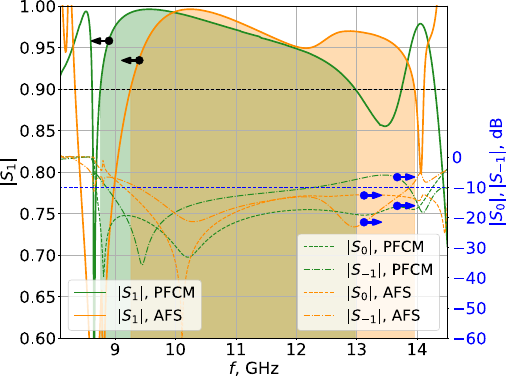}
\caption{Numerically calculated frequency spectra of the scattering parameters $S_1$, $S_{0}, S_{-1}$ (into the propagating FB modes with indices $n=1,0,-1$ accordingly) from perfectly conducting corrugated anomalous reflectors designed for $\theta=60^{\circ}$, $f_0 = 10$~GHz using both methods. For clarity, the operational frequency bands of reflectors designed using the PFCM and AFS methods are highlighted with green and orange filling, respectively. The black dashed line indicates the level defining the operational frequency band, while the blue dashed line refers to the level of $-10$~dB for the spurious FB modes.}
\label{Fig_5_Frequency_properties_example}
\end{figure}

The numerically evaluated frequency behavior of the scattering parameters $S_1$, $S_{0}, S_{-1}$ for the same examples as for Fig.~\ref{Fig_4_Syntesis_examples} is illustrated in Fig.~\ref{Fig_5_Frequency_properties_example}. The level of 0.9 is displayed with a black dashed line. We define $f_{\text{max}}$ and $f_{\text{min}}$ as the upper and lower bounds of the band $\Delta f$, respectively. In Fig.~\ref{Fig_5_Frequency_properties_example} the operational bands of the reflectors synthesized using the PFCM and AFS methods for $\theta=60^{\circ}$ are highlighted with green and orange filling, respectively. It can be checked that the corresponding spurious scattering levels $|S_{0}|, |S_{-1}|$ hold below the level of $-10$~dB, which is represented in Fig.~\ref{Fig_5_Frequency_properties_example} with the blue dashed line. For the given examples shown in Fig.~\ref{Fig_5_Frequency_properties_example} the relative operational bandwidth $\Delta f / f_0$ is $42.2\%$ for the PFCM method and $47.0\%$ for the AFS method. Although the difference between bandwidths is small, the upper and lower frequency bounds for the AFS and PFCM methods differ substantially, namely $13.9$ vs. $12.9$~GHz, and $9.3$ vs. $8.8$~GHz, accordingly.
\begin{figure}[h!]
\centering
\includegraphics{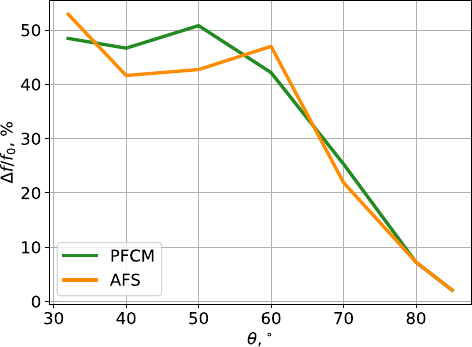}
\caption{The comparison of the relative operational bandwidth of the anomalous MS reflectors synthesized using the PFCM and AFS methods for different anomalous reflection angles $\theta$.}
\label{Fig_6_Comparison_of_frequency_properties}
\end{figure}

Now, let us compare the results obtained with both methods for different reflection angles $\theta$ ranging from $32^{\circ}$ to $85^{\circ}$. Upon examination of the AFS method solutions with the criterion based on \eqref{Numerical_criterion_for_purity}, it can be revealed that with increasing $\theta$, the number of solutions $M$ having higher $\zeta$ than for the PFCM method decreases significantly, as represented in Table \ref{Table_satisfyed_solutions_for_AFS}. Interestingly, for end-fire anomalous reflection, that is, for $\theta \ge 85^{\circ}$ no AFS solution can realize a better purity of the reflected wave front compared to the PFCM method.
\begin{table}
\caption{\label{Table_satisfyed_solutions_for_AFS}%
Number $M$ out of 16 AFS solutions, corresponding to different number of evanescent FB modes $N_l \in [0,15]$, exceeding the reflected wave front purity of the direct solution by the PFCM method for different desired $\theta$.}

\begin{tabular}{>{\centering\arraybackslash\ensuremath}p{0.8cm} >{\centering\arraybackslash\ensuremath}p{0.69cm} >{\centering\arraybackslash\ensuremath}p{0.69cm} >{\centering\arraybackslash\ensuremath}p{0.69cm} >{\centering\arraybackslash\ensuremath}p{0.69cm} >{\centering\arraybackslash\ensuremath}p{0.69cm} >{\centering\arraybackslash\ensuremath}p{0.69cm} >{\centering\arraybackslash\ensuremath}p{0.69cm}}
\hline
\hline
$\theta$ & $32^{\circ}$ & $40^{\circ}$ & $50^{\circ}$ & $60^{\circ}$ & $70^{\circ}$ & $80^{\circ}$ & $85^{\circ}$ \\
\hline
$M$ & $13\hphantom{1}$ & $11\hphantom{1}$ & $10\hphantom{1}$ & $6\hphantom{1}$ & $4\hphantom{1}$ & $1\hphantom{1}$ & $0\hphantom{1}$ \\
\hline
\hline
\end{tabular}
\end{table}

The relative operational bandwidths for the direct solution using the PFCM method and for the best selected solution using the AFS method are compared in Fig.~\ref{Fig_6_Comparison_of_frequency_properties} versus the desired anomalous reflection angle. The greatest difference is achieved at reflection angles smaller than $60^{\circ}$ (e.g., at $\theta=50^{\circ}$ the difference achieves $8.1\%$, where the PFCM method appears to be more preferable). In contrast, for $\theta>60^{\circ}$ the difference in the realized bandwidth in both methods is imperceptible.

\begin{figure*}
\includegraphics{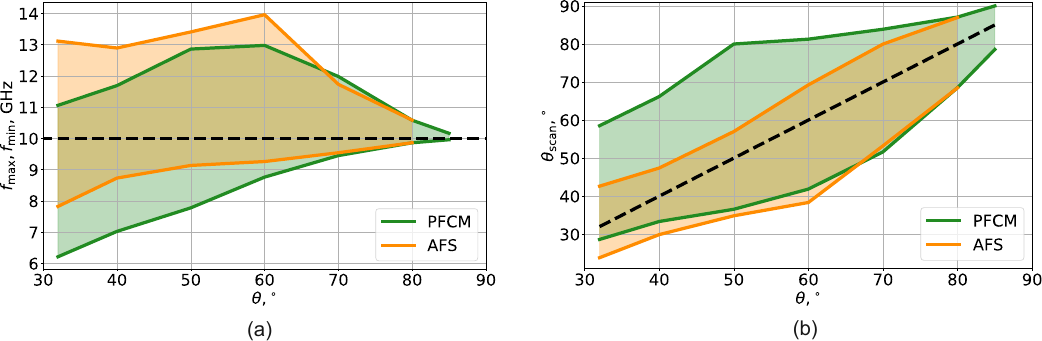}
\centering
\caption{\label{Fig_7_Frequency_scaninng_capability}Comparison of the PFCM and AFS synthesis methods in terms of frequency bandwidth and frequency scan angles depending on the synthesis angle $\theta$ of anomalous reflection at $f_0=10$~GHz (green lines and filling correspond to the PFCM method, orange lines and filling correspond to the AFS method): (a) frequency bounds of the operational band $f_{\text{min}}, f_{\text{max}}$ (the black dashed line represents the frequency at which the reflector is synthesized); (b) the possible range of frequency scan angles $\theta_{\text{scan}}$ (the dashed black line $\theta_{\text{scan}} = \theta$ refers to reflection to the desired angle 
$\theta$ at the frequency $f_0$ of synthesis).}
\end{figure*}
However, there is a notable difference between the two methods, which consists in different dependences of $f_{\text{max}}$ and $f_{\text{min}}$ on $\theta$, as shown in Fig.~\ref{Fig_7_Frequency_scaninng_capability}(a).
The upper bound frequency $f_{\text{max}}$ for the AFS method is on average $1.2$~ GHz higher than for the PFCM method in the range of reflection angles from $32^{\circ}$ to $60^{\circ}$, and similarly for $f_{\text{min}}$. For $\theta>60^{\circ}$, the difference between the two methods in terms of the frequency position of the upper and lower bounds of the band vanishes. This result has a considerable effect on the frequency scan capability of the anomalously reflected beam. Indeed, for the anomalously reflected FB mode with index $n=1$ the reflected beam angle $\theta_{\text{scan}}$ at frequency $f$ different from the goal frequency $f_0$, at which the reflector is designed, can be found as
\begin{equation}
    \theta_{\text{scan}} = \arcsin \left( \dfrac{f_0}{f} \, \sin \theta \right).
    \label{Frequency_scan_angle}
\end{equation}
According to \eqref{Frequency_scan_angle} different frequencies within the operational frequency band correspond to different scan angles.

In Fig.~\ref{Fig_7_Frequency_scaninng_capability}(b) the variation of $\theta_{\text{scan}}$ within the entire band $\Delta f$ for every required anomalous reflection angle $\theta$ at $f=f_0$ is shown for both methods. For clarity, the ranges of possible scan angles of the PFCM and AFS methods are highlighted by green and orange filling, respectively. The black dashed line represents the case of $f = f_0$, that is, $\theta_{\text{scan}} = \theta$. Since $f_{\text{max}}$ takes larger values in the case of the AFS method for $\theta \in [32^{\circ},60^{\circ}]$, that method provides by $1.9^{\circ}-4.8^{\circ}$ broader frequency scan angular range close to the normal direction of the MS (i.e., z-axis) compared to the PFCM method. In contrast, $f_{\text{min}}$ takes smaller values in the case of the PFCM method for $\theta \in [32^{\circ},60^{\circ}]$, therefore the PFCM method provides considerably broader range of frequency scan angles near the grazing reflection direction than the AFS method (the benefit reaches $23^{\circ}$).

\section{Conclusion}

The two most important synthesis methods for anomalous reflectors based on local impedance MSs have been compared in terms of the operational bandwidths and angular range of the frequency scan, i.e., the PFCM and AFS methods. The comparison was done by parametrically repeating the analytical synthesis and numerical simulations of perfectly curved corrugated structures (curved for the PFCM method and flat for the AFS method) with the same discretization. 

In conclusion, both methods were found to provide a comparable relative bandwidth. Moreover, for $\theta>60^{\circ}$, both methods exhibit an identical and relatively large bandwidth of anomalous reflectors implemented using corrugated structures. The bandwidth shrinks similarly for both methods as $\theta$ increases to grazing angles.

However, the PFCM method was found to be more suitable for the frequency scan of the anomalously reflected beam near the grazing angles with the benefit of $23^{\circ}$ in the scan angular range over the AFS method. At the same time, the AFS method provides better conditions for the frequency scan near the normal direction, showing scan ranges broader by $1.9^{\circ}-4.8^{\circ}$ for the the goal anomalous reflection angles $\theta \in [32^{\circ},60^{\circ}]$. 
This difference can be explained by the different positions of the operational frequency band relative to the goal frequency of synthesis. 

The above comparison also shows that the computational complexity of reaching the same purity of the anomalously reflected wave front for the AFS and PFCM methods is different. While the PFCM method requires solving one nonlinear differential equation leading to a direct solution, the AFS method requires solving multiple nonlinear optimization problems for a number of excitation surface waves to be determined, which is much more time-consuming. However, the PFCM method requires a higher fabrication complexity because of a curved MS profile and cannot provide the low-profile structure, which is desirable in many applications.

The investigated properties of both methods should be considered in designing relatively broadband anomalous reflectors based on all-metal corrugated structures operating in the centimeter and millimeter-wave ranges.

\section{Acknowledgment}
This work was supported by the Ministry of Science and
Higher Education of the Russian Federation (Project No. 075-15-2022-1120).

\nocite{*}
\bibliographystyle{unsrt}
\bibliography{references.bib}

\end{document}